# TEMPERATURSKALAN OCH BOLTZMANNS KONSTANT

G.J. Ehnholm och M. Krusius*
Aalto Universitet, Esbo, Finland

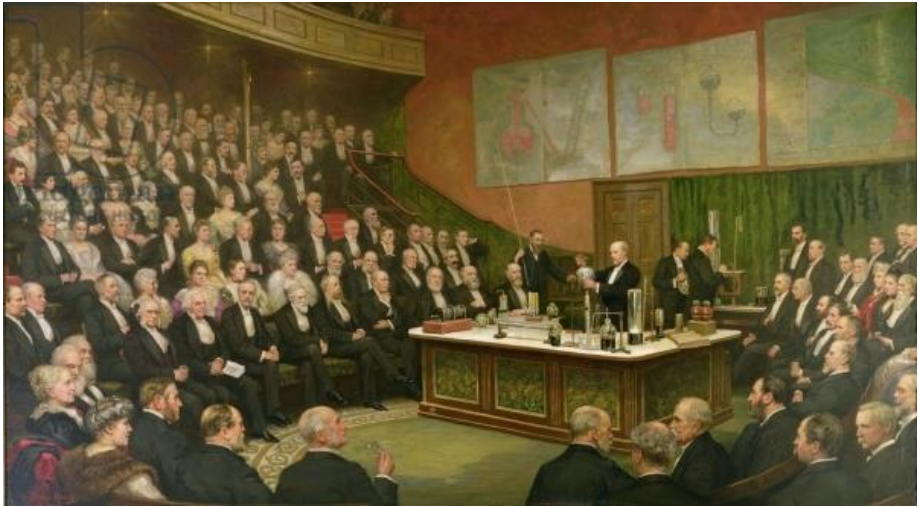

Tavlan *" föreläsningen av Sir James Dewar över vätskeformig vete en fredag kväll 1904 vid Royal Institution i London"* är målad av Henry Jamyn Brooks.

Det nyaste systemet av mått och massa grundar sig på naturkonstanterna, som skall vara sinsemellan kompatibla. Ett exempel är Boltzmanns konstant $k_B$, som anger tätheten av den termiska energin $k_B T$. För att uttrycka basenheten $T$, absolut temperaturen i kelvin, behöver man en internationell överenskommelse för temperaturskalan. Skalan har definierats med hjälp av fixpunkter, som är temperaturer för olika fastransformationer. Speciellt viktig har varit trippelpunkten av vatten vid 273,16 K. Fixpunkt temperaturerna fastställer den *internationella temperaturskalan ITS* inom *Si-systemet (Système international d'unités)*. Temperaturmätningen grundar sig på fysikaliska lagar och på egenskaper hos lämpliga termometriska material som har valts för att fastslå temperaturskalan. För att bestämma Boltzmanns konstant har nya precisionsmetoder utvecklats under de senaste årtiondena. Exempel är olika arter av gastermometri, som baser på den allmänna gaslagen, samt termiska bruset av elektriska laddningsbärare. Med dessa medel har det blivit möjligt att fixera värdet av Boltzmanns konstant med en relativ osäkerhet av $\Delta k_B / k_B \lesssim 10^{-6}$ (≙ 1 ppm). Från och med år 2019 ersatte det fixerade värdet $k_B$ = 1.380 649 × 10⁻²³ J/K definitionen av en Kelvin grad.

**Historia**

Människan har sedan urminnes tider sysslat med mätning och måttenheter. Är det lika kallt eller varmt idag som igår? Temperaturen är en viktig storhet som reglerar vårt dagliga liv, därför behöver vi en termometer som kan visa temperatur, till exempel genom höjden på en kvicksilverpelare eller läget på en visare. Resultatet kan beskrivas med ord, som kallt, varmt, hett. En mera exakt tolkning är att ge resultatet som ett tal, på samma sätt som t.ex. för vikt och längd. Detta kräver en överenskommelse om en skala. Därför skapades *Système international d'unités* (SI- systemet), vars ursprung härstammar från den franska revolutionen: Då gjorde man en prototyp t.ex. för vikt. Med hjälp av denna kunde man fastställa att man hade fått rätt mängd bröd av torghandlaren. De exakta värdena för storheterna har sedan dess kontinuerligt preciserats.

Antagligen gjordes de första dagliga meteorologiska observationerna, som berodde på reella mätningar, under Medici-tiden i Toscana från 1654 till 1667 [1]. Enligt Galileo Galilei hade han upptäckt ca. 1593 hur man konstruerar ett termoskop, med vilket man kunde jämföra vilket av två objekt var varmare, men som inte var lämpligt för nyttiga kvantitativa observationer. Den första verkliga termometern, som inte samtidigt var en barometer och som hade färgad alkohol i ett slutet glasrör, sägs ha utvecklats 1654 av storhertigen Ferdinando II de' Medici di Toscana (Fig. 1).

Temperaturmätning med exakta och jämförbara värden blev till 1714 när tysken Daniel Fahrenheit utvecklade kvicksilvertermometern med en fast skala som han sedermera 1724 standardiserade till den form som vi nu känner som Fahrenheit-skalan. Celsiusskalan med "centigraderingen" infördes litet senare 1743 av den svenska astronomen Anders Celsius. Den använder två fixpunkter eller fastransformationer, fryspunkten 0°C och kokpunkten 100°C av rent vatten under normalt tryck. Celsiusskalan är praktisk för dagligt bruk, men i fysikens värld är det den absoluta temperaturen som gäller.

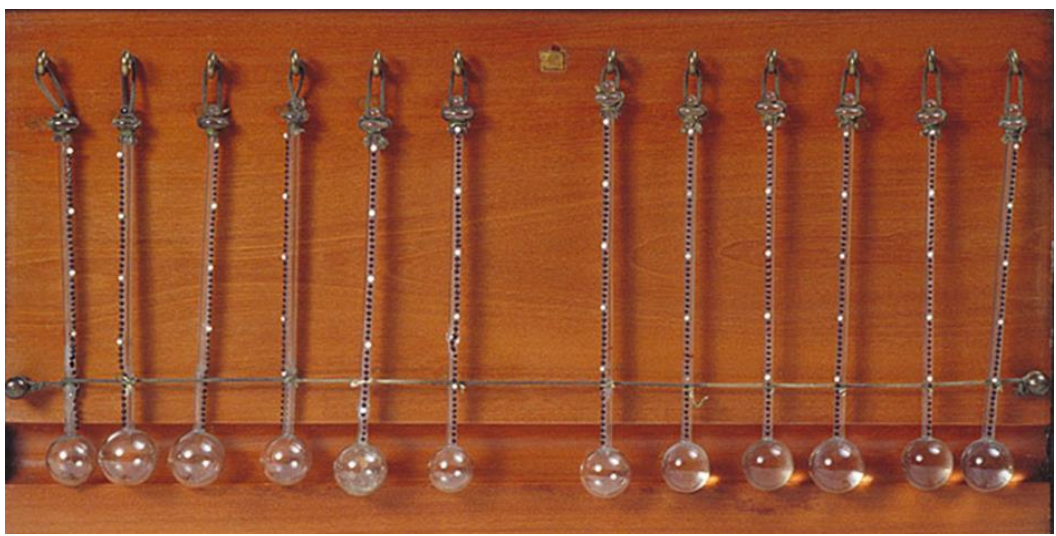

**Fig. 1.** Termometrar från andra hälften av 1600-talet i Galilei Museums samlingar i Firenze. Avståndet mellan de vita fläckarna motsvarar ungefär 10 grads temperaturskillnad medan hos svarta fläckarna är det en tiondel av detta.

Förhållandet mellan den praktiska temperaturskalan och den absoluta temperaturen blev uppenbarligen känd rätt sent. Begreppet den absoluta nollpunkten dök upp som ett vetenskapligt forskningsresultat redan 1695 när fransmannen Guillaume Amontons konstaterade att extrapolationen av utslaget för hans luftgastermometer pekade på att dess volym skulle gå till noll vid tillräcklig avkylning, en temperatur som skulle motsvara nu för tiden ungefär -240 °C. Denna observation bekräftades under senare år av många forskare, bland annat Jacques Charles under 1780-talet och Joseph Gay-Lussac 1802. Men argumenteringen verkade osäker för man antog, att även om luften förhöll sig som en så kallad "permanent gas" vid dåtida laboratorieförhållanden, så var det troligt att den inte skulle fortsätta sitt elastiska tillstånd till den extrapolerade nollpunkten.

Ett stort inflytande i diskussionen hade William Thomson [2], som i sin utredning 1848 över temperaturen förklarade dess samband med värme, som absorberas av materien, och me-kaniskt arbete, såsom man utövar för att lyfta ett kilogram till en meters höjd. Hans syn på den absoluta temperaturen fann växande godkännande, i synnerhet som han under sina senare år blev en av de mest betydande brittiska fysikerna, bättre känd med sitt adlade namn Lord Kelvin. Numera placeras den absoluta nollpunkten till -273,15 °C.

**Tävlan mot absoluta nollpunkten**

Under första hälften av 1800-talet upptäckte man att många olika gaser övergick i vätskeform då man kylde dem. Detta påvisade bland andra den brittiska fysikern och universalsnillet Michael Faraday. Men det fanns också många gaser som han inte kunde kondensera med då kända metoder, genom att kyla dem i ett bad av eter under högt tryck. Till dem hörde vanliga gaser som kolmonoxid (som kräver -78 °C), metan (-164 °C), syre (-183 °C), kväve (-196 °C) och väte (-253 °C). Dessa kände man som "permanenta gaser".

Vid denna tid experimenterade James Joule med en termodynamisk process som nu bär hans namn ("Joule expansion") och som William Thomson blev intresserad av. Samma funktionsprincip används för kylning i många kylskåp. År 1852 utarbetade Thomson en termodynamisk förklaring för kylningseffekten och förklarade att expansionsmetoden kunde man använda för att kondensera permanenta gaser. Kännedomen av Joule-Thomson-kylning gav upphov till lågtemperaturfysiken och tidsperioden blev senare känd som "tävlan mot den absoluta nollpunkten".

1877 lyckades Louis Cailletet, och oavhängigt Raoul Pictet, tillverka en dimma av små droppar syre. Deras metod var att låta förkyld gas under 300 atm tryck expandera genom en smal ventil. Genom att förbättra tekniken och apparaturen blev det sedan 1883 möjligt för Zygmunt von Wroblewski och Karol Olszewski att producera syre i vätskeform i ett provrör. Samma teknik kunde de sedan använda för kväve och kolmonoxid, men vätet behöll alltjämt sitt epitet som permanent gas. Ännu 1895 gjorde Olszewski i Krakow ett försök, men förgäves. Det var den holländske teoretikern Johannes van der Waals som i sin doktorsavhandling 1873 framlade den första modellen för tillståndsekvationen av en reell gas med interpartikel växelverkan (Box 2). Med hjälp av van der Waals ekvation kan man visa att det för varje gas finns en maximal så kallad inversionstemperatur, som anger den översta gränsen för temperaturen där JT-processen resulterar i kylning. För kväve är gränsen 350 °C, men för väte blir JT-kylningsmetoden mera svårtillgänglig då det krävs en utgångstemperatur på -71 °C.

Olszelwskis konkurrent i tävlan mot nollpunkten var James Dewar i London. Han hade infört användningen av vakuumisolerade termosflaskor för att bevara och hantera gaser i vätskeform. I lågtemperaturlaboratoriet kallas ett sådant kärl ännu idag "dewar". År 1898, utrustad med en sådan blev det möjligt för Dewar att kondensera väte, vilket han gjorde även i ett berömt möte av Royal Society 1904 med medlemmarna som åskådare [3]. Kokpunkten av väte är 20 K och redan samma år nådde Dewar 14 K, där väte fryser till fast form. Det gjorde han genom att minska ångtrycket inom gasfasen ovanför vätskan med en vakuumpump.

En parallell utveckling var upptäckten av ädelgaserna, där utforskningen i första hand gjordes av den brittiska fysikern William Ramsay. År 1869 hade man hittat en ny absorptionslinje i solens spektrum som motsvarade ett obekant element och fick namnet helium. 1895 lyckades Ramsay isolera helium gas från mineralen pechblände där den ansamlas genom radioaktivt sönderfall som alfapartiklar. Övriga ädelgaser destillerades så småningom från flytande luft eller naturgas från olika källor. Nu gällde det att bestämma deras kok- och fryspunkter.

Helium var ett särskilt fall: Olszewski försökte kondensera helium, men lyckades inte. Men inte heller Dewar, han gjorde flera försök under följande år, men det visade sig att hans tekniska beredskap inte var tillräckligt utvecklad för dessa krävande undersökningar. Det var först Heike Kamerlingh Onnes, en professor och kollega till van der Waals vid Leiden Universitetet, som bemästrade de tekniska kraven och 1908 kondenserade helium gasen till vätska vid 4,2 K och normalt tryck. Genom att minska ångtrycket över heliumvätskan med en vakuumpump kylde han den till ungefär 1,5 K som blev lägsta temperaturrekordet inom ca. tre decennier. Med dessa prestationer inledde han en ny epok i den kryogeniska tekniken och framförallt i lågtemperaturfysiken. Det var Kamerlingh Onnes själv som lade grundstenarna genom att upptäcka supraledningsfenomenet och de första indikationerna av det supraflytande tillståndet, de viktigaste och totalt oväntade företeelserna i det nya temperaturområdet man uppnått. För upptäckten av heliumvätskan belönades han med Nobelpriset 1913. Ett betydligt lägre temperaturrekord av 0,25 K nåddes först 1933 då W.F. Giauque och D.P. Mac-Dougall införde en ny kylningsmetod som kallas adiabatisk demagnetisering.

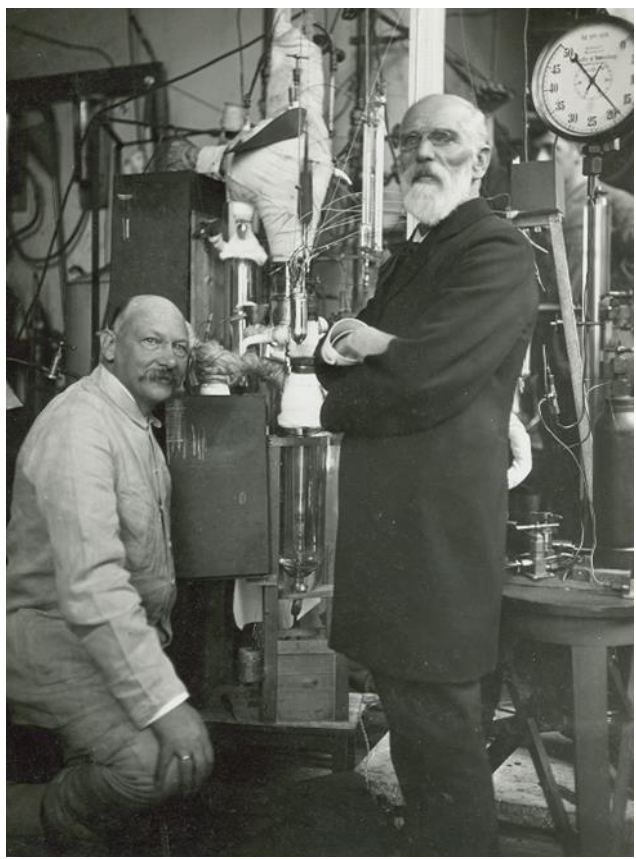

**Fig. 2.** Professorerna Kamerlingh Onnes och Johannes van der Waals i lågtemperatur laboratoriet vid Leiden universitetet. Året är 1913 när Kamerlingh Onnes belönades med Nobelpriset för sina forskningar vid låga temperaturer som hade lett till förvätskandet av helium. Kollegan van der Waals hade fått sitt Nobelpris 1910 för utvecklingen av tillståndsekvationen för gaser och fluider.

**Den ständiga processen att revidera temperaturskalan**

Sedan 1889 har det funnits ett internationellt behov att göra överenskommelser om hur bland annat temperaturmätningen skall ske och vilken noggrannhet måste uppnås. Forskare och byråkrater från olika länder har församlats i konferenser, som kallas *Conference Generale des Poids et Mesures (CGPM)* och organiseras av *Bureau Internationale des Poids et Mesures (BIPM)* i Paris. I dessa möten har gamla ordningar reviderats, nya konventioner har införts som motsvarar mera avancerade tekniska beredskap, och rekommendationer har förberetts för forskare om nya behov för kalibrering.

Bland dessa forskningsuppdrag är ett viktigt och alltid aktuellt område den internationella temperaturskalan. Ofta är det inte temperaturen i sig som det gäller utan de processer som behövs för standardisering av mätningsinstrument eller mätningsprocesser och deras kalibrering. Målsättningen är att mätningarna följer standardiserade överenskommelser och är reproducerbara var och när som helst. Fig. 3 specificerar typiska åtgärder som överenskommelserna har omfattat med åren.

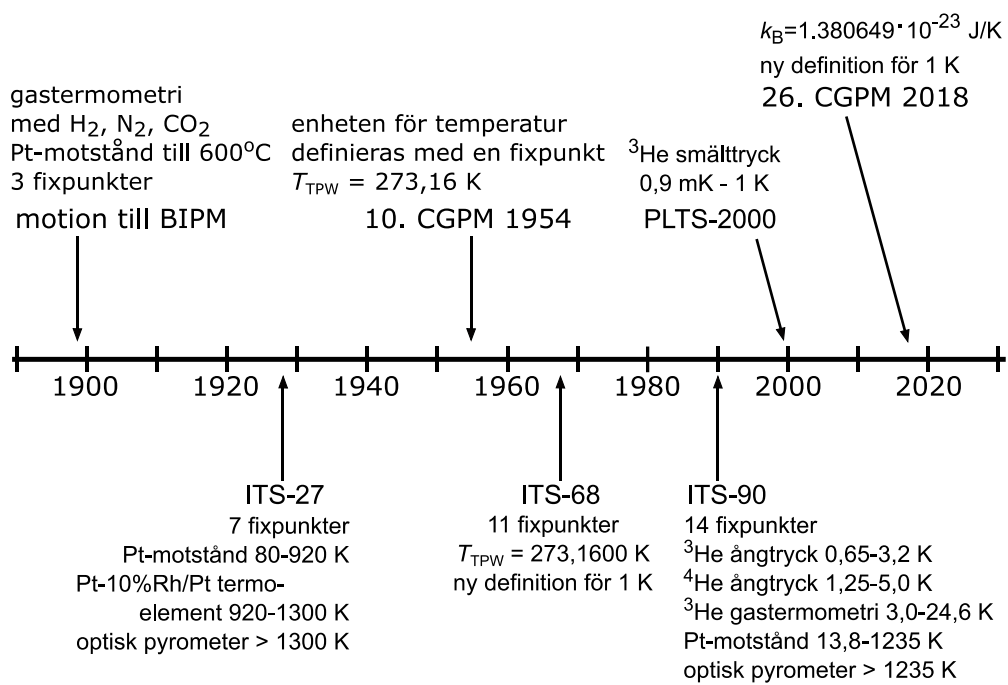

**Fig. 3.** Tidslinjen av de viktigaste åtgärderna för att fixera den internationella temperaturskalan *ITS* som överses av *Conference Generale des Poids et Mesures (CGPM).* Skalan är baserad på internationella överenskommelser om vilka fixpunkter används för kalibrering och deras värden, samt mätningsmetoderna och analytiska formuleringar för en slät och kontinuerlig interpolation av mätvärdet som funktion av temperaturen.

Bland rekommendationerna för temperaturkalibrering är t.ex. beslutet från 1954 att den viktigaste termodynamiska fixpunkten, trippelpunkten $T_{TPW}$ av rent vatten, ligger exakt vid 273,1600 K. Denna fixpunkt används för kalibrering och också som den temperatur, där värdet av Boltzmanns konstant skulle bestämmas. Samtidigt har denna överenskommelse gett en ny definition för en Kelvin. Definitionen "rent vatten" fick också en ny betydelse i resolutionen från 2005, där kraven för vattnets isotopiska komposition klarlagdes. När det krävs allt större noggrannhet, då blir specificeringarna mera ingående och komplicerade.

### SI-systemet

Strävandet att standardisera temperaturmätningen är bara ett av uppdragen inom SI-systemet vars utveckling överses av BIPM. Fig. 4 visar några tidpunkter då banbrytande reformer har tagits i bruk inom SI-systemet genom internationella avtal om förhållanden och värden mellan de fysikaliska basenheterna. Den ursprungliga tankegången vid slutet av 1700-talet var att jämföra basenheterna med naturvärden, såsom till exempel att 1 kg är lika med mas-san av en liter vatten. För att kontrollera noggrannheten måste temperaturen, lufttrycket, litermåttet, vattnets renhet osv. specificeras. Dessa kvalificeringar antyder var svårigheterna ligger hos en sådan definition.

Under franska revolutionen införde man därför en speciellt framställd prototyp för kilogrammassan, en artefakt som var en cylinder av platina och som uppbevarades i Paris. Problemet nu var att om man ville ha exakt visshet av kilogrammassan sa måste man resa till Paris och jämföra sitt eget duplikat av massan med den parisiska prototypen. I de mest avancerade försök av en sådan jämförelse har man uppnått en otrolig hög noggrannhet av litet bättre än $10^{-7}$, men som ändå inte räcker för dagens krav.

Den snabba tekniska utvecklingen efter andra världskriget gjorde det möjligt att sammanbinda basenheterna med naturvärden genom noggranna tekniska mätningar. Speciellt frekvensmätningen blev viktig. Det första exemplet var måttet för en meter som 1960 definierades med hjälp av den orange-färgade strålningen av krypton-86 med en våglängd av ca 606 nm, som är enkelt att åstadkomma med en gasurladdningslampa. En ännu mera pålitlig spektrallinje var hyperfina övergången i caesium-133 atomens grund tillstånd, $\Delta \nu_{Cs}$ = 9,1926... GHz, som är väl skyddad från utvärtes magnetiska och elektriska fält. 1964 bestämdes $^{133}$Cs linjen för en prefererad frekvensstandard och 1967 övertogs den för att fastslå storheten av en sekund: en sekund är $9,1926 \cdot 10^9$ perioder av $^{133}$Cs hyperfina strålning.

På 1970-talet konstaterade man att laserljuset av många olika molekylära absorptionslinjer kunde stabiliseras med en relativ noggrannhet av bättre än $10^{-8}$. 1975 slog man sedan fast ljushastigheten i vakuum med en noggrannhet av 9 siffror till $c$ = 2,9979...$10^9$ m/s. Med hjälp av detta värde kom man 1983 till den slutsatsen, att man kunde ge meter-längden en ny definition: en meter är den längd som ljuset framskrider i $1/2,9979 \cdot 10^9$ sekund. Det betyder, att om man till exempel har $^{133}$Cs linjen $\Delta \nu_{Cs}$ = 9,1926... GHz för fogande, så blir våglängden $\lambda_{Cs}$ = $c/\Delta \nu_{Cs}$ = (2,9979·10$^8$ m/s) / (9,1926·10$^9$ 1/s) = 0,0326 m, eller en meter motsvarar 30,6633189 våglängder. Det här beslutet att deklarera ljushastigheten $c$ som den första naturkonstanten med ett fixerat värde, $c$ =299 792 458 m/s, blev ett banbrytande steg i SI-systemets utveckling under dess sekellånga historia. Fig. 4 ger en tidsmässig översikt av dessa framsteg.

Under de senaste åren har SI-systemet sålunda genomgått en djupgående förändring, som slutfördes i det 26. mötet av CGPM 2018: i stället för de sju basenheterna meter, sekund, kilogram, amper, kelvin, mol och candela slog man fast värden på sju fysikaliska naturvärden. Utöver frekvensen av $^{133}$Cs hyperfina övergång och ljushastigheten har fem andra naturkonstanter fixerats: Plancks konstant $h$ (= 6.626 × 10$^{-34}$ J s), elementarladdningen $e$ (= 1.602 × 10$^{-19}$ C), Boltzmanns konstant $k_B$ (= 1.380 × 10$^{-23}$ J/K), Avogadros konstant $N_A$ (= 6.022 × 10$^{23}$ mol$^{-1}$), och ljusekvivalenten för monokromatisk strålning vid 540 × 10$^{12}$ Hz frekvens $Kcd$ (= 683 lm/W). För det mesta har dessa värden blivit uppmätta i åtminstone tre olika experiment med en noggrannhet av 3 ppm eller bättre.

Naturkonstanterna skall vara jämförbara och kompatibla, för de ingår i fysikens lagar. Till exempel energirelationerna bör uppfyllas, dvs. $E = mc^2 = k_B T = h\nu = eV$. På så sätt har SI-systemet befriat sig från artificiella prototyper som förut fixerade basenheterna. Härmed övergavs också den sista prototyp-baserade definition, som symboliserades av platina-iridiumcylindern i Paris, urkilogrammet från 1879, välkänd som internationell prototyp för kilogram [4]. Den väsentliga förändringen från SI-systemets synpunkt är att basenheternas definition inte mera är det dominerande innehållet, utan genom att fixera värdet av Boltzmanns konstant $k_B$ till exempel, ger vi definitionen av en Kelvin en ny mening. Det gör den internationella temperaturskalan ITS inte överflödig, det betyder att både existerande och nya fixpunkter måste kanske justeras att motsvara den nya definitionen.

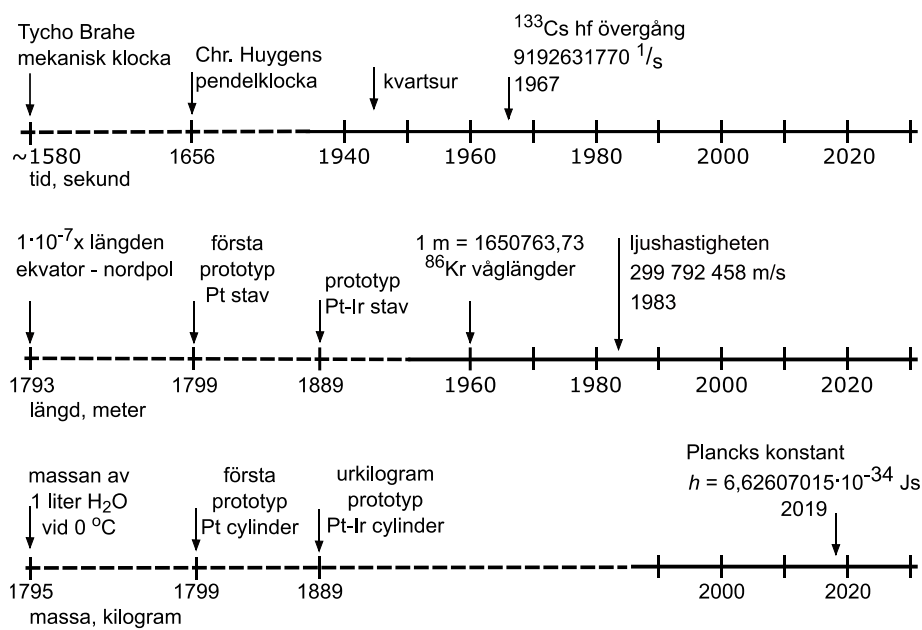

**Fig. 4.** Illustration av tidslinjen efter franska revolutionen som sammanfattar de viktigaste åtgärderna i den utveckling som lett från mekaniska standardmått till ett system bestämt genom värden på naturkonstanter. Utvecklingen av tre av SI-systemets sju basenheter visas. Dessa är tid, längd, och massa uttryckt i enheterna sekund, meter, och kilogram, som antogs ursprungligen 1889 som basenheter för *MKS-systemet*, i motsats till det tidigare *cgs-systemet* som härstammar från mitten av 1800-talet.

**Gastermometri**

En enkel illustration av skillnaden mellan den gamla och nya konventionen kan ges med hjälp av allmänna gaslagen, som antagligen representerar det första sammanhanget där Boltzmanns konstant dyker upp i skolan. Gaslagen anger tryck *p*, volym *V* och temperatur *T* för en gas utan interpartikel växelverkan:

$$pV = NRT. \qquad (1)$$

*N* är mängden av gas i mol och $R = N_A k_B$ den *allmänna gaskonstanten*. I stället för mol kan vi lika väl uttrycka gasmängden med antalet partiklar $n = NN_A$ och gaslagen med hjälp av Boltzmanns konstant: $pV = n k_B T$.

Man kan använda gaslagen för gastermometri och bestämma produkten $k_B T$, om man gör tryckmätningar vid konstant volym. Om man därtill bestämmer temperaturen, får man ett värde för $k_B$. Detta var en mätningsmetod för $k_B$ ända till 1970-talet. I praktiken är experimentet svårt att utföra med tillräcklig noggrannhet och man har övergått till bättre alternativ.

Men säg att vi skulle använda i gaslagen i stället för en kelvin t.ex. $a \cdot$ kelvin som basenhet. Om ekvationen ännu skall hålla, så måste Boltzmanns konstant få ett nytt värde $k_B/a$. Det här var konventionen före 2019. Numera är det Boltzmanns konstant som har ett fixerat värde. Om vi kallar värdet $k_B{}^*$, så måste basenheten för kelvin multipliceras med $k_B/k_B{}^*$, dvs. basenheten får ett ändrat värde. Det låter självklart, men ändå blir det komplicerad när det handlar om kalibreringsarbete med osäkerheter på ppm nivån och otaliga fixpunkter som man har bestämt under tidens lopp med varierande noggrannhet.

**Boltzmanns konstant $k_B$**

Österrikaren Ludwig Boltzmann (1844 – 1906) var en av 1800-talets främsta fysiker. Otaliga ekvationer, formuleringar och begrepp bär hans namn. En av de mest kända är ekvationen för entropin (1877), som skapade sambandet mellan den makroskopiska termodynamiken och statistiska fysiken och därmed senare med kvantfysiken. Det här uttrycket för entropin innehåller en konstant faktor med entropins dimension, som Boltzmann själv antagligen inte specificerade. Det tycks ha varit Max Planck som först gav värdet och namnet till Boltzmanns konstant i sitt arbete över svarta kroppens strålning (1901).

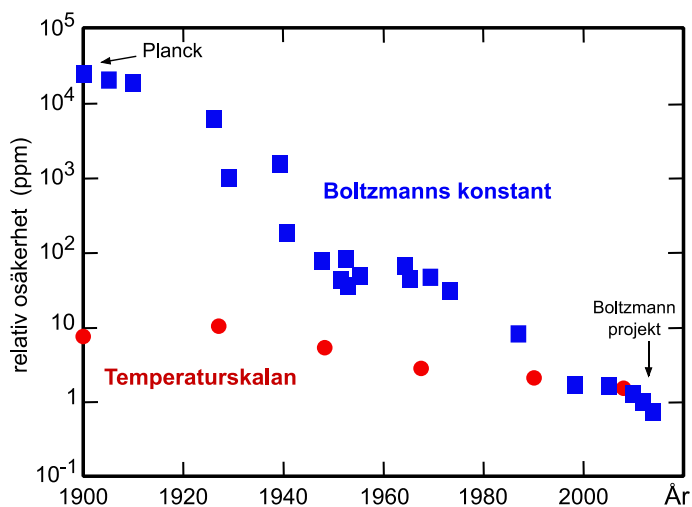

**Fig. 5.** Den 120 åriga Boltzmanns konstanten. En schematisk illustrering av den relativa osäkerheten i värdet av Boltzmanns konstant, sedan den infördes av Max Planck 1901, och den samtidiga utvecklingen i fastställandet av en kelvin.

Idag är Boltzmanns konstant en av SI-systemets hörnstenar, sedan resolutionen från 2018 blev godkänt att inkludera den till en av de sju naturkonstanter med fixerat värde: $k_B = 1.380649 \times 10^{-23}$ J/K. Visserligen har värdet kunnat bestämmas bara med en relativ noggrannhet av ca 1 ppm, vilket är drygt två storleksordningar sämre än hos Plancks konstant $h$ eller Avogadros tal $N_A$. Under de två senaste decennier har ett 30-tal publikationer diskuterat och presenterat resultat för mätningen av $k_B$. Precisionsmätningar har gjorts i internationellt samarbete, bland annat den välkända Boltzmann kollaborationen [5], och i 11 publikationer har det rapporterats en precision av $\pm$ 4 ppm eller bättre.

Det har visat sig att vara ytterst svårt att minska osäkerhetsmarginalen av $\pm$ 1 ppm med dagens tekniska möjligheter. Denna omständighet visar att precisionsmätningen av $k_B$ är en komplicerad och mångsidig process, jämförbar med temperaturmätningen själv, för det är tills vidare bara mätningen av vattnets trippelpunkt som har utförts med en precision av ca 1 ppm. Det är både temperaturstabiliseringen och -homogeniteten som är krävande. Det finns tills vidare endast några få metoder som lämpar sig för precisionsmätningar av $k_B$ och bara med tre av dem har det hittills åstadkommits en tillräcklig noggrannhet: mätning av ljudhastigheten och av dielektriska susceptibiliteten i ädelgaser, antingen He eller Ar, och Johnson-Nyquist brus i en metallisk ledning.

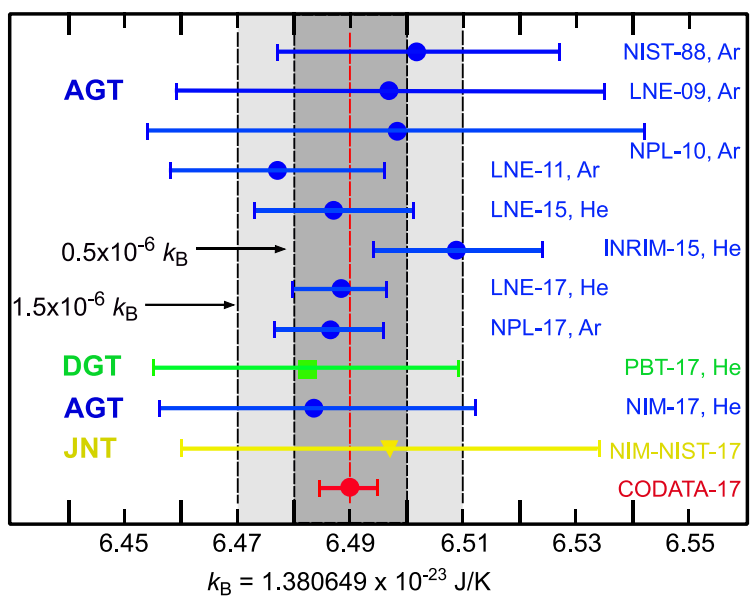

**Fig. 6.** Precisionsbestämningar av Boltzmanns konstant och dess osäkerhetsmarginal under de senaste trettio åren. Utvecklingen med tre olika metoder listas med tidsförflyttning uppifrån neråt (AGT – akustisk gastermometri, DGT – dielektrisk gastermometri, och JNT – brustermometri). De olika mätningarna märks av förkortningen för institutet (som är alla nationella metrologiska institut), året för publicering av resultatet, och den använda gasen. Horisontala axeln visar de tre sista decimalerna i $k_B$ = 1,380649×10$^{-23}$ J/K.

**Akustik gastermometri**

**Fig. 7**. Klotformad kopparresonator [6] av 3 l volym och 14 kg massa med god värmeledningsförmåga för ökad temperaturhomogenität och mekanisk stabilitet. Resonatorn befinner sig inom en termostat och två värmeskydd för temperaturstabilisering. Resonatorn är fylld med $^4$He gas vid $T_{TPW}$ = 273,16 K temperatur och används för att mäta både akustiska resonanser vid 2 – 10 kHz frekvens och mikrovågresonanser vid 2 – 10 GHz.

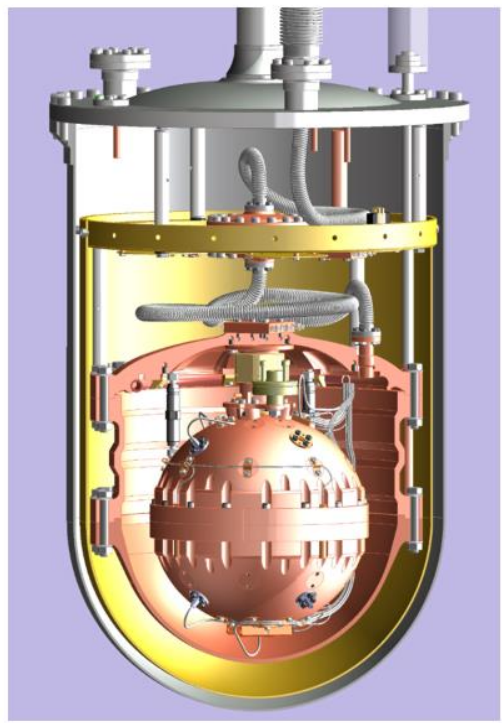

Tillsvidare har den mest framgångsrika tekniken för att bestämma värdet av $k_B$ varit akustisk gastermometri, som har gett de noggrannaste resultaten. Tekniken har utvecklats från ljudhastighetsmätningar för gaser under de fyra senaste årtionden. Man får värdet för produkten $k_B T_{TPW}$ genom att bestämma ljudhastigheten från resonansvillkoret för en akustisk resonator [5]. I det senaste skedet har man valt en klotformad resonator (Fig. 7), som är fylld med He eller Ar gas. I dessa precisionsmätningar har det blivit möjligt att sänka osäkerheterna i $k_B$ till ppm nivån [6].

Den massiva kopparresonatorn ligger inbyggd i en termostat med effektiva värmeskydd med temperaturen stabiliserad vid $T_{TPW}$. Resonanstillståndet bestäms från vågekvationen, som i den sfäriska geometrin leder till en lösning i form av en produkt av Bessel och sfärisk-harmoniska funktioner [5]. För mätningen används en ren radialt symmetrisk lösning (med l = 0), där Bessel funktionens $J_0(k\,r) = 0$ rötter kallas $z_{0n}$. I praktiken registrerar man i mätningarna flera radiala resonanser n, för att minska osäkerheter genom redundanskontroll. Låt oss för enkelhets skull negligera resonansordningen n och uttrycka vågvektorn k med hjälp av frekvensen f samt ljudhastigheten c:  $k = 2\pi f/c$. Därmed är resonansfrekvensen

$$f_0 = \frac{c\,z_0}{2\pi a} \quad , \tag{2}$$

där resonatorns radie a kan uttryckas med dess volym V som $a = (3\,V/4\pi)^{1/3}$. Om vi placerar här uttrycket för ljudhastigheten från Box 1, så får vi en ekvation, som man använder i mätningarna för att bestämma Boltzmanns konstant $k_B$,

$$k_B = \frac{c^2\,m}{\gamma T} = m\left(\frac{f_0}{z_0}\right)^2 \frac{\left(6\pi^2 V\right)^{2/3}}{\gamma T} \quad . \tag{3}$$

Här är $\gamma = C_p/C_v = 5/3$ relationen mellan värmekapaciteterna för en monoatomisk gas och m atomvikten.

Först och främst bestämmer man alltså resonansfrekvensen $f_0$, som är oberoende av gastrycket i första storleksordning. Växelverkan mellan gasatomerna leder dock till ett svagt inflytande av trycket som ger upphov till korrektionstermer och som kan tas i beaktande genom de första två eller tre virialkoefficienterna. Virialexpansionen av allmänna gaslagen är den enklaste metoden att inkludera växelverkan mellan gasatomerna i första approximation (Box 2). Här finns det två möjligheter: antingen litar man på teoretiska *ab initio* kalkylationer för virialkoefficienterna, som numera är mycket noggranna speciellt för en enkel atom som He, eller mätningarna görs som funktion av trycket och extrapoleras till p → 0. Genom att anpassa extrapolationen till en polynom av tryck bestämmer man i själva verket virialkoefficienterna experimentellt. I praktiken använder man bägge metoder för att öka redundans och noggrannhet.

De första ljudhastighetsmätningarna med en resonator, som då hade cylinderform, gjorde man vid slutet av 1970-talet. Sedan dessa mätningar, som nådde en relativ noggrannhet av ±25 ppm, har det varit en stor utmaning under de fyra senaste decennier att minska osäkerheterna till en nivå av 1 ppm. Till exempel, hur skall man bestämma den kritiska längden a eller volym V? Mätningarna i den cylinderformiga resonatorn på 1970-talet gjordes vid en fixerad frekvens av 5.6 kHz genom att justera längden a för att uppfylla resonanskonditionen och sedan mäta längden med optisk interferometri. Vid slutet av 1980-talet bestämde man volymen av den sfäriska resonatorn genom att fylla den med kvicksilver som man sedan vägde. Slutligen på 2010-talet började man använda den sfäriska resonatorn samtidigt för resonansmätningar med mikrovågor, där ljushastigheten har sitt accepterade värde och genomsnittsradien a fås från resonansvillkoret lika som i ekvationen (2).

Akustisk gastermometri är en intensiv mätning, vilket betyder att som första storleksordning är resultatet oberoende av gasmängden. Man behöver till exempel inte mäta gasmängden i det adsorberade ytlagret på resonatorväggen. Trots detta kommer den största korrektionen till de obehandlade mätvärdena av den akustiska resonansen från detta ytlager, som förorsakar en termoakustisk förskjutning av resonansfrekvensen och en ökning av resonansbredden. Gastätheten i ytlagret minskar med distansen från kopparväggen med en exponentiell dämpningslängd av ~ 50 μm som är proportionell till kvadratroten av termiska diffusiviteten $\lambda/(\rho C_p)$, där $\lambda$ är värmeledningskoefficienten, $\rho$ tätheten och $C_p$ värmekapaciteten i gasen. Ytlagret ger en kontribution till frekvensförskjutningen $\Delta f/f$ som, beroende på gastrycket, motsvarar 40 – 400 ppm och som man tar i beaktande med hjälp av analytiska kalkylationer, där det behövs ett värde för $\lambda$ från *ab initio kalkyler*.

Enligt ekvation (3) skall gasen vara så ren som möjligt för att undgå resonansförskjutningar. De mest noggranna resultaten har man fått med $^4$He gas, som förs till resonatorn under ett tryck av 1 – 10 bar. I ren $^4$He gas är den enda betydande orenheten $^3$He isotopen. Med kryogeniska metoder kan man minska $^3$He koncentrationen till en nivå som inte mera påverkar resonansmätningen. Däremot har desorption av restgaser (speciellt $H_2O$) från kopparmaterialet en mätbar effekt. Den kan minskas genom att skölja resonatorns insida kontinuerligt med ett konstant reglerat flöde av renad $^4$He gas. Trots detta påverkar desorptionen från kopparväggarna slutresultatet på en nivå av ~ 0.1 ppm.

Stabiliteten av temperaturen och dess fördelning är en ytterligare osäkerhetskälla. Temperaturmätningen utförs med flera Pt-resistanstermometrar som är fördelade över kopparklotet. Men här är frågan inte osäkerheten i själva temperaturmätningen, utan hur pålitligt och nära till $T_{TPW}$ temperaturen i apparaturen kan stabiliseras och hur homogen temperaturfördelningen i resonatorn blir.

Som vi ser, består osäkerhetsbudgeten av många diverse andelar som man har lyckats minska genom omsorgsfulla åtgärder. Metrologiska precisionsmätningar är överhuvudtaget ytterst arbets- och tidskrävande, en kontinuerlig kamp för att öka noggrannheten (Fig. 4). I akustisk gastermometri fördelar sig osäkerheterna först och främst mellan mätningen av temperaturen och den akustiska resonansen, medan mindre andelar till osäkerhetsbudgeten kommer från bestämmandet av resonatorns dimensioner samt av gasrenheten och gasmängden. För att fixera värdet av $k_B$ fordras det, att den relativa osäkerheten av alla dessa andelar tillsammans inte överskrider 1 ppm. De mest avancerade mätningarna av ett forskarkonsortium, som kallar sig *Boltzmann projekt* (LNE-CNAM) [6], har nu uppnått det här målet.

Akustisk gastermometri har utvecklats till den mätteknik, där man har kunnat förbättra noggrannheten med rimliga insatser. Man har uppnått en noggrannhetsnivå $\Delta k_B/k_B \sim 10^{-6}$ där det är svårt att förbättra ytterligare med en eller två storleksordningar. En stor fördel är en enkel fysikalisk basis, där mindre osäkerhetskällor kan kompletteras med teoretisk analys och *ab initio* kalkylationer av gasegenskaperna. Resonansmätningar vid både akustiska och mikrovåg frekvenser med samma apparatur i samma förhållanden ökar kontrollmöjligheterna, speciellt då man kan använda många olika resonansmoder i båda fallen. Det ger fler chanser att undvika störningar, t.ex. vid frekvenser som råkar ligga nära mekaniska resonanser i resonatorns kopparskal. Dessa effekter växer med ökat gastryck och därför är det praktiskt från teknisk synpunkt att mätningarna kan utföras som funktion av trycket och sedan extrapoleras till nolltryck. Med nutida tekniska lösningar kan man använda akustisk gastermometri över ett brett temperaturområde, ungefär 5 – 500 K.

**Dielektrisk gastermometri**

Kapacitansmätningar av den dielektriska konstanten för $^4$He som funktion av gastrycket *p* var den andra tekniken som man började utveckla till en precisionsmätning av $k_B$ [7]. Med åren visade det sig att detta initiativ växte till ett verkligt "tour de force" uppdrag: Tre storheter måste mätas samtidigt med en noggrannhet bättre än 1 ppm, nämligen temperaturen vid $T_{TPW}$, kapacitanstillägget *C*(*p*) – *C*(0), när gastrycket ökas från vakuum till *p*, och som tredje storhet själva gastrycket *p*. Särskilt kapacitans- och tryckmätningen vid 1 ppm nivå fordrade långvarigt utvecklingsarbete.

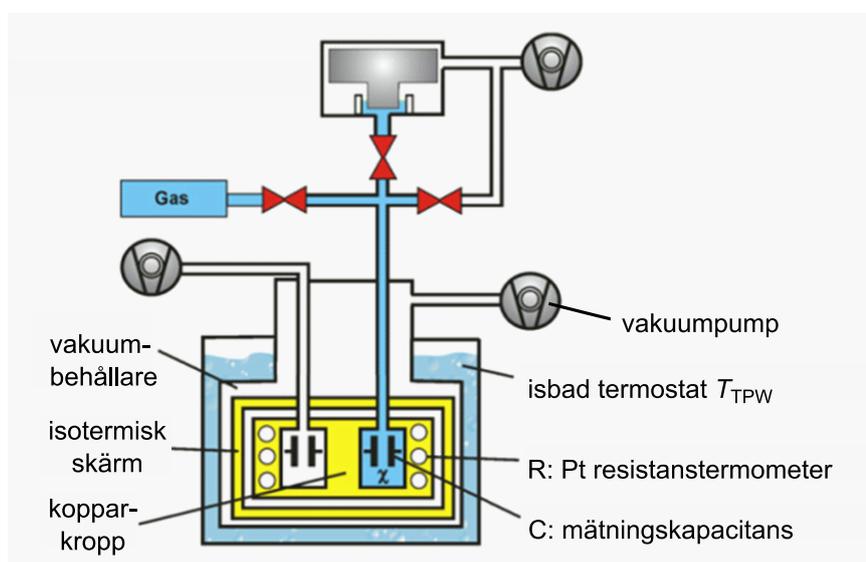

**Fig. 8.** Schematisk bild av den kapacitiva mätinrättningen [7] för heliums dielektriska susceptibilitet $\varepsilon_r$ - 1. Referenskondensatorn i vakuum ligger på vänster och mätkondensatorn fylld med He gas vid tryck *p* på höger. Kondensatorerna är inbäddade i en termostat vid konstant temperatur $T_{TPW}$, som observeras med flera platina resistanstermometrar. Alla dessa delar ligger inom en vakuumbehållare. Tryckmätningen sker med en klassisk mekanisk balansvåg.

Dielektrisk gastermometri utgår från Clausius-Mossotti ekvationen. Den upptäcktes av Ottaviano Mossotti (1850) och Rudolf Clausius (1879) oberoende av varandra, för att bestämma dielektriska konstanten $\varepsilon = \varepsilon_r \varepsilon_0$ av ett material med en molekylär elektrisk polariserbarhet $\alpha_0$ och en täthet av *n* molekyler per volymenhet,

$$\frac{\varepsilon_r - 1}{\varepsilon_r + 2} = \frac{\alpha_0 n}{3 \varepsilon_0} \,, \qquad (4)$$

där gastätheten *n* bestäms från gastrycket *p* genom idealgaslagen: $n = p/(k_B T)$. Ekvationen håller särskilt väl för en ädelgas med ringa växelverkan. När det gäller en enkel atom som $^4$He kan den elektriska dipolpolariserbarheten $\alpha_0$ bestämmas från *ab initio* kalkylationer med en relativ osäkerhet av ∼ 10$^{-7}$. I mätningarna bestämmer man alltså från den relativa kapacitansförändringen $[C(p) - C(0)]/C(0) = \varepsilon_r$ - 1 den relativa dielektriska konstanten $\varepsilon_r$ samt trycket *p*. Då har man alla värden för att bestämma Boltzmanns konstant från ekvation (4).

Vanligen använder man en cylindrisk kondensator, där $^4$He gasen befinner sig mellan elektroderna och storleksordningen av totalkapacitansen är ∼ 10 pF. Då kapacitansförändringen måste mätas med en relativ noggrannhet på 0,1 ppm nivån, använder man höga tryck ända till 70 bar och kompletterar idealgaslagen med extra korrektionstermer från virialexpansionen (Box 2). En särskild svårighet vid så höga tryck är då kravet att bestämma därtill den effektiva kompressabiliteten $\kappa_{eff}$ av kondensatorkonstruktionen, även om kondensatorn är framställd av rostfritt stål eller volframkarbid. Därför dyker en extra korrektionsterm upp i kapacitansförändringen $[C(p) - C(0)]/C(0) = \varepsilon_r - 1 + \varepsilon_r \kappa_{eff} p$, som är linjär med tryck, men som måste bestämmas från mätningarna med en relativ osäkerhet av ∼ 10$^{-4}$.

Mätningen av dielektriska konstanten har dock också fördelar: gasmängden behöver icke mätas och helium gasens isotopiska renhet är inte väsentlig. Men heliums renhet med hänsyn till andra gaser är ytterst viktig och kontrolleras kontinuerligt med en masspektrometer. Den mest problematiska orenheten i kondensatorgasen är vatten som har en två storleksordningar större molekylär elektrisk polarisation $\alpha_0$ än helium och vars ursprung i kondensatorn är desorption från byggmaterialet. Sammantaget liknar undersökningen av osäkerhetsbudgeten för mätning av dielektriska konstanten lika invecklade granskningar som i fall av akustisk gastermometri. För att fastslå värdet av $k_B$ för det nya SI-systemet fordrades det att åtminstone två olika mätmetoder används med överlappande relativ standard osäkerhet mindre än 3 ppm. Detta krav blev fylld 2017 när det långvariga projektet vid Physikalisch-Technische Bundesanstalt i Berlin (PTB) fick osäkerhetsbudgeten av sina dielektriska mätningar nedskuren till $\Delta k_B / k_B \sim 2 \times 10^{-6}$.

**Brustermometri**

För omdefinieringen av en kelvin krävdes det att åtminstone två olika mätmetoder ger samma resultat med relativ överlappande osäkerhet som är mindre än 3 ppm. Det blev ett stort evenemang 2017 när noggrannheten också i brustermometrin kom att uppfylla 3 ppm kravet. Då var det klart att resultaten med tre olika mätningsmetoder gjorde det möjligt att fixera värdet av $k_B$. I alla tre fall hade utvecklingsarbetet för att förbättra noggrannheten av mätmetoden krävt flera decennier.

I motsats till de olika former av primära gastermometriska metoder, framför allt den akustiska och dielektriska gastermometrin, där avvikelserna från idealgasen på grund av reella gasers egenskaper blir viktiga, beror brustermometrin på en ren elektronisk mätning av den termiska brusspänningen i ett elektriskt motstånd, som är termiskt förankrad till den temperatur som skall mätas. Den fysikaliska grunden är tillförlitlig, men tekniskt har uppnåendet av den önskade osäkerhetsmarginalen krävt mödosamt arbete.

I brustermometri mäter man den termiska Johnson brusspänningen över en resistans *R*. Bruset ger den absoluta temperaturen *T* enligt Nyquists formel. Termiska bruset uppstår som följd av ledningselektronernas värmerörelse, som ger upphov till ström- och spänningsvariationer i ett elektriskt motstånd, som sedan också syns som spänningsskillnader mellan motståndets terminaler. Medelvärdet av strömvariationerna är noll, en användbar effekt går alltså inte att få. Men man mäter i stället medelvärdet på kvadraten av variationerna, d.v.s. brus-effekten. Denna är oberoende av frekvensen och kallas därför vitt brus, analogt med vitt ljus där olika frekvenser är inblandade med samma amplitud. För temperaturer nära 300 K och frekvenser under 1 GHz beskrivs den genomsnittliga kvadratiska spänningen $\langle V_T^2 \rangle$ i Johnson brus av Nyquists lag,

$$\langle V_T^2 \rangle = 4k_B TR \, \Delta f , \qquad (5)$$

med en relativ noggrannhet bättre än $10^{-9}$ ($\triangleq$1 ppb). Här är $\Delta f$ den bandbredd över vilken bruset mäts. I praktiken är den termiska brussignalen ofta svag jämfört med allmänt förekommande störningskällor. Trots detta har man använt brustermometri för att bestämma $k_B$ till en relativ osäkerhet på endast 10 ppm under en integrationsperiod av ungefär ett dygn. För de bästa resultaten har man gått ända upp till integrationsperioder på 100 dygn.

I allmänhet försöker man minimera det termiska bruset som något icke önskvärt i elektroniska kretsar. Men i mätningarna av Boltzmanns konstant gäller det att maximera bruset genom att öka bandbredden och resistansen av bruskällan. Men redan att bestämma värdet av motståndet R och bandbredden $\Delta f$ i ekvationen (5) med tillräcklig noggrannhet bereder svårigheter. I praktiken görs mätningarna med hjälp av analog till digital konverterare som ger signalerna digitalt som en tidsserie. Mätningarna konverteras sedan till en funktion av frekvensen med hjälp av en dator, som räknar om dem med Fourier transformation. På så sätt kan bandbredden $\Delta f$ definieras exakt.

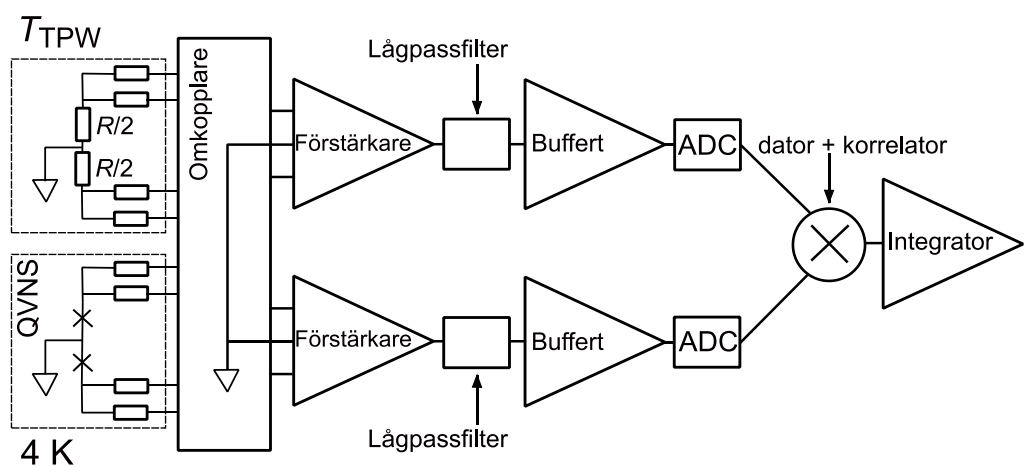

**Fig. 9.** Tekniska realiseringen av brustermometri för $k_B$ mätningen [8] - en schematisk illustrering av mätningskretsen. Korrelationstekniken, med en omkopplare vid ingångskretsen för korskoppling mellan två signalkällor och två förstärkarkedjor, har utvecklats för metrologiska låg-brusmätningar för att korrigera instabilitet i förstärkningen och extra brus från förstärkarkedjan. Korskorrelation, signalmultiplikation, medelvärdering, och bandbreddsdefinition utförs digitalt med dator.

Den mest sofistikerade utvecklingen har varit att jämföra signalen från bruskällan till en kalibrerad referens, en syntetisk spänningskälla med samma värde av impedans. I de senaste brusmätningarna [8] har referenskällan varit en matris av Josephson junktioner, en kvantiserad spänningskälla för vitt brus, som kallas QVNS (för quantized voltage noise source). Båda bruskällor kopplar man med en fyra-tråds anslutning till mätelektroniken (Fig. 9), vilket gör det möjligt att definiera motståndsvärden mera exakt. Med en omkopplare ansluter man båda bruskällor omväxlande till vardera av de två parallella signalvägar. På så sätt kan man eliminera drift i förstärkningen och korrelerade störningar i ledningarna och i förstärkarna, om man utför korskorrelation av signalerna senare.

För detta ändamål konverteras signalerna efter förstärkning och filtrering till digitalform, dvs. till två tidsserier som är synkrona med omkopplaren. Dessa förs till en dator med vilken korskorrelationen genomförs. I slutskedet för signalvärderingen bestämmer man kvoten av signalerna från bruskällan och QVNS referensen. Tekniska skäl begränsar tillsvidare resistansvärdet av QVNS referenskällan till bara 200 $\Omega$. Samtidigt måste båda källor fungera som rena resistanser med ett flatt spektrum som funktion av frekvensen. I praktiken tycks detta krav begränsa maximala frekvensen till under några hundra kHz. Som bruskälla använts två plåtformiga motstånd med 100 $\Omega$ resistans, gjorda av Ni-Cr metallegering och termiskt stabiliserade vid $T_{TPW}$ temperaturen i ett is bad.

I den mest avancerade mätningen [8] nåddes en noggrannhet av $\Delta k_B/k_B \sim 2{,}7 \times 10^{-6}$. Siffran är huvudsakligen begränsad av mätningsstatistiken. Som funktion av längden av mätningstiden $\tau_m$ förhåller sig den relativa osäkerheten i brustermometri som

$$\frac{\langle \Delta V_T^2 \rangle}{\langle V_T^2 \rangle} \geq \frac{1}{\tau_m \, \Delta f} \quad , \tag{6}$$

om alla andra variablerna kan hållas konstanta. För att minska osäkerheten under 1 ppm borde den nuvarande integrationsperioden $\tau_m$ = 100 dygn höjas till ett års längd, vilket stöter på stora problem. Elektromagnetisk interferens, speciellt magnetiska störningar vid låga frekvenser, måste elimineras. Därför utförs dessa mätningar i ett elektriskt avskärmat rum beläget i ett underjordiskt laboratorium för att minska mekaniskt buller och hålla avstånd till alla slags källor som genererar elektromagnetiska störningar.

Dessa precisionsmätningar för bestämningen av Boltzmanns konstant gjordes vid den relativ höga $T_{TPW}$ temperaturen. De lägsta temperaturer där man har gjort kalibrering av temperaturskalan med brustermometri sträcker ända ned till 40 $\mu$K. Mot höga temperaturer växer brussignalen, vilket gör termometrin lättare och tekniskt möjlig kanske ända till 1500 C. Sålunda har brustermometrin det allra bredaste användningsområdet av alla termometriska metoder, 8 storleksordningar. Troligen kommer brustermometrin vara ett av de viktigaste verktygen för att realisera ett kommande avtal av den internationella praktiska temperaturskalan, som kommer att bära namnet IPTS202X (Fig. 3).

**Kalibrering**

Genom att fixera värdet av Boltzmanns konstant $k_B$ har kalibreringen av temperaturskalan fått ett nytt innehåll. Om vi tar som exempel brustermometrin och ekvationen (5), så beror bruseffekten på produkten $k_B T$. En mätning av $\langle V_T^2 \rangle$ ger då värdet av denna produkt, men om man kan använda $k_B$ med ett fixerat värde, får man direkt temperaturen $T$. Fig. 10 visar ett exempel av ett sådant experiment [10].

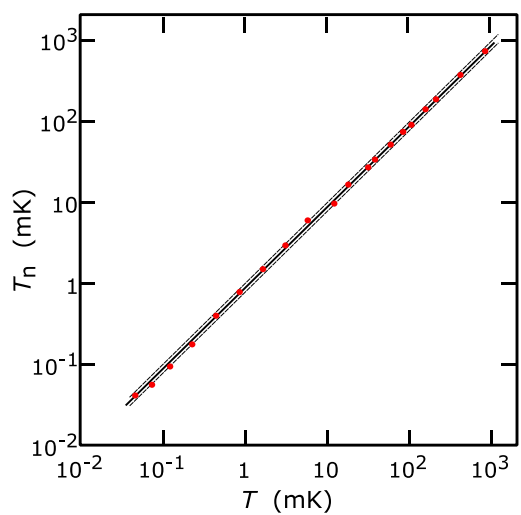

**Fig. 10.** Kalibrering av brustermometri vid de lägsta temperaturerna [10]. Brustemperaturen $T_n$ är inritad på vertikal axeln som funktion av den samtidiga mätningen av en referenstemperatur $T$ på horisontal axeln. Referenstemperaturen kommer från en kärnmagnetisk mätning av platinas kärnmagnetiska susceptibilitet ($T$ < 12 mK) eller från resistansvärdet av ett $RuO_2$ motstånd ($T$ > 12 mK). Den kontinuerliga linjen motsvarar en perfekt kompatibilitet av brustemperaturen och referensvärdet. Streckade linjerna avgränsar en osäkerhetsmarginal av $\pm$ 5 %.

På ordinatan i Fig. 10 finns brustemperaturen som man har bestämt från den induktiva signalen av virvelströmmar, förorsakade av elektronernas termiska motion i en kopparcylinder. Abskissaxeln visar temperaturen antingen från den kärnmagnetiska susceptibiliteten av platina ($T$ < 12 mK) eller temperaturen av ett $RuO_2$ motstånd ($T$ > 12 mK). Den kärnmagnetiska susceptibiliteten följer Curies lag, där susceptibiliteten $\chi \propto 1/T$. Susceptibilitettermometern är därför en primär termometer, som behöver kalibreras bara vid en känd temperatur. $RuO_2$ motståndet är däremot en sekundär termometer, som måste kalibreras punktvis mot en fixpunkttermometer. De kalibrerade motståndsvärdena anpassar man till en kontinuerlig polynomartad funktion av temperatur, som sedan används för att översätta motståndsvärden till temperaturer.

Som vi ser i Fig. 10, är jämförelsen av brustemperaturen till kalibreringsvärdena god, alla mätpunkter ligger inom $\pm$ 5 % marginaler av det gemensamma beroendet. I princip skulle man ha kunnat använda ordinatan från dessa mätningar för att bestämma värdet av $k_B$ i stället att bara demonstrera ett linjärt beroende. Tyvärr är en vanlig kalibreringsmätning, som här i Fig. 10, inte tillräckligt noggrann: Boltzmanns konstant skulle komma med en osäkerhet på några procent i stället för några ppm!

Detta illustrerar varför det är så viktigt att bestämma $k_B$ med den bästa möjliga noggrannhet som kan uppnås. En ny kalibrering, som i Fig. 10, måste fastslås vid åtminstone en fixpunkt som tillhör IPTS protokollet. Lägsta temperaturen i Fig. 10 är $\sim 10^{-5}$ K medan den viktigaste fixpunkten $T_{TPW}$ är $\sim 10^2$ K. Arbetsmängden som ingår i att härleda temperaturer över 7 storleksordningar gör att det är i praktiken inte möjligt. Hittills har man gjort det genom att använda olika fixpunkter som är spridda med jämna mellanrum över temperaturskalan. Dessa har man bestämt med olika noggrannhetsmarginaler under hela senaste seklet. I praktiken utnyttjar man vanligen den närmaste fixpunkten som har fastslagits i tidigare forskningsarbete, men som inte nödvändigtvis är kompatibel med $T_{TPW}$ fixpunkten med en noggrannhet av storleksordning $\sim$ 1 ppm.

Nu, när $k_B$ har fått ett fixerat värde, som med $\sim$ 1 ppm noggrannhet motsvarar fixpunkten $T_{TPW}$, kan man använda det i alla kommande kalibreringar och referera på så sätt till fixpunkten $T_{TPW}$. Till exempel i Fig. 10 kan man då avläsa temperaturen direkt från ordinatan. Figuren föreställer då en jämförelse mellan tre olika termometrar som visar ett linjärt ömsesidigt förhållande, där alla tre termometrar är sinsemellan kompatibla inom en osäkerhetsmarginal av $\pm$ 5 %.

Denna revolutionära förändring i metrologin av temperaturskalan grundar sig på ett hängivet utvecklingsarbete av en stor mängd forskare under flera årtionden med de tre ovanbeskrivna mätningsmetoderna som nu har öppnat en ny framtid i metrologin.

**Resumé**

Ända tills 2019 använde man som definition av en kelvin trippelpunkten av vatten vid $T_{TPW}$ = 273,16 K, där den relativa osäkerheten var $4 \times 10^{-7}$. Sedan dess innebär den nya definitionen att en kelvin är den förändringen i den termodynamiska temperaturen $T$ som motsvarar ett tillskott i termisk energi $k_B T$ av $1.380\,649 \times 10^{-23}$ joule. En relativ noggrannhet av ca. $10^{-6}$ har uppnåtts i precisionsmätningar av skalfaktorn, Boltzmanns konstant $k_B$, med tre olika metoder: Mätningar av ljudhastigheten i helium eller argon gas, mätning av den dielektriska konstanten av helium gas, och mätningar av Johnson-Nyquist brus i ett elektriskt motstånd.

För metrologin är den nya kelvindefinitionen av stor betydelse. Den överensstämmer fullständigt med den tidigare fixpunkts definitionen av trippelpunkten för rent vatten vid $T_{TPW}$ = 273,16 K. Men ännu viktigare är att den grundläggande standarden, som man jämför till i kalibreringsmätningar, inte längre är värdet av vattenstandarden $T_{TPW}$, utan temperaturen kan nu kalibreras vid såväl låga som höga värden med den optimalaste metoden och kan sedan refereras till Boltzmanns konstant.

En djupare konsekvens av den nya definitionen blir uppenbarlig från vetenskapshistorien, som på ett övertygande sätt visar, att ny fysik uppstår när nya mätningar med extrem precision tas i bruk. Därför väcker den nya definitionen och förbättrade noggrannhetsnivån stora förväntningar för temperaturmätningen i framtidens forskning.

\*   E-post: <matti.krusius@aalto.fi>

-----------------------------------------------------------

**Faktruta 1. Ljudhastigheten i gas**

Gastermometrin grundar sig på idealgaslagen och i akustiska gastermometrin mäter man ljudhastigheten. Ljudet framskrider som en vågrörelse av variationer i gasens kompression. Då ljudhastigheten är flera hundra meter per sekund, är täthetsvariationerna adiabatiska. I vågekvationen för gastäthetens $\rho$ möter man ljudhastigheten $c$ som en koefficient $c^2 = (\partial p/\partial \rho)_S$. Med hjälp av termodynamiska potentialer kan man uttrycka den i en mera brukbar form. Likaväl kan vi överta här från kinetiska gasteorin ljudhastigheten för en Maxwell-Boltzmann gas

$$c^2 = \frac{1}{3}\gamma \langle v^2 \rangle \; , \tag{1}$$

var $\gamma$ är den adiabatiska expansionsfaktorn, $\gamma = C_p/C_v = 5/3$, relationen mellan värmekapaciteterna för en monoatomisk gas. Den kvadratiska medelhastigheten av gasatomerna $\langle v^2 \rangle$ kan uttryckas enligt ekvipartitionsekvationen för kinetiska energin med

$$\frac{1}{2} m \langle v^2 \rangle = 3 \times \frac{1}{2} k_B T \; .$$

Således skrivs ljudhastigheten som

$$c^2 = \frac{1}{m}\gamma\, k_B T = \frac{\gamma R T}{M} \; , \tag{2}$$

där $M = m\, N_A$ är molvikten. Uttrycket med mätvärdet proportionell till temperaturen, $c^2 \propto T$, gör att man kan använda ljudhastigheten som *primär termometer*.

En beaktansvärd egenskap hos ekvationen (2) är att i första storleksordning är ljudhastigheten oberoende av gastrycket. Därför lämpar sig mätningen av ljudhastigheten särskilt väl för att bestämma temperaturen. Men om växelverkan mellan gasatomerna tas i beaktande, då måste man inkludera de första koefficienterna från virialexpansionen (Faktruta 2) till ljudhastigheten och i stället av ekvation (2) skriva:

$$c^2 = \frac{1}{m}\gamma\, k_B T + A_1\, p + A_2\, p^2 + \ldots \; . \tag{3}$$

För heliumatomer är, jämfört med $c^2$, storleksordningen av virialkoefficienterna $A_1 \sim 10^{-4}$ och $A_2 \sim 10^{-6}$.

-----------------------------------------------------------

**Faktruta 2. Virial koefficienterna**

I termodynamikens historiska utveckling har den enkla idealgaslagen spelat en stor roll. Ändå blev det redan tidigt klart, att när temperaturen sänks och närmar sig kondensation av en reell gas blir växelverkan mellan gasatomerna påtagliga och idealgaslagen inexakt. Virialexpansionen var en av de första metoderna att inkludera korrektionstermer i gaslagen.

Antagligen var det Kamerlingh Onnes 1902 som först införde korrektioner i form av en serieutveckling som funktion av partikeltätheten. Man kan antingen bestämma de första koefficienterna från mätningar eller beräkna deras värden från kalkylationer baserade på mikroskopiska modeller. Kalkylen utgår från en beräkning av växelverkan mellan två atomer som funktion av deras avstånd. Första virialkoefficienten $A_1$ beror på den attraktiva två-partikelväxelverkan, medan $A_2$ innebär både två- och tre-partikelväxelverkan. Den enklaste modellen som tillåter en uppskattning av koefficienterna $A_1$ och $A_2$ är den så kallade van der Waals gasen för vilken Johannes van der Waals presenterade de första analyserna i sin doktorsavhandling från år 1873.

Den enklaste modellen för växelverkan $U(r)$ mellan två gas atomer som funktion av deras avstånd $r$ består av den så kallade hard-core repulsionen för distanser $r < r_o$, var atomerna är inte tillåtna, $U(r) \to \infty$, och av van der Waals attraktionen för $r > r_o$, som har kort räckvidd med ett brant distansberoende av $U(r) = - u_o(r/r_o)^6$.

Virialexpansionen är tillståndsekvationen för virialgasen i serieform,

$$\frac{p}{nk_BT} = 1 + A_1 n + A_2 n^2 + \{n^3\} \quad , \tag{1}$$

där koefficienterna $A_i$ kan uträknas t.ex. med hard-core modellen.

Enligt van der Waals skrivs tillståndsekvationen för en mol gas i följande form:

$$p = \frac{RT}{V_m - b} - \frac{a}{V_m^2} \quad , \tag{2}$$

var $V_m = V/N$ är mol volymen och $N$ antalet mol. Konstanten $a$ representerar gaspartiklarnas attraktion, som förorsakar en extra reducering av gastrycket. Konstanten $b$ i sin tur härleds från hard-core repulsionen: $b$ representerar den andel av mol volymen som är utesluten från det som finns kvar för termiska rörelsen: därmed är den proportionell till hard-core volymen = $(4/3) \pi r_o^3$. Om man utvecklar $(1 - bn)^{-1}$ i ekvation (2) som funktion av partikeltätheten $n = N/V$ till binomserie, så får van der Waals ekvationen samma form som virialexpansionen (1),

$$\frac{p}{nk_BT} = 1 + (b - \frac{a}{RT}) n + \frac{1}{2} b^2 n^2 + \{n^3\} \quad . \tag{3}$$

Virial koefficienterna för van der Waals gasen är således $A_1 = b - a/(RT)$ och $A_2 = b^2/2$. Virialkoefficienter för olika gaser, som har bestämts genom att anpassa till experimentella resultat, hittar man i tabeller.

-----------------------------------------------------------

**Faktruta 3. Johnson-Nyquist brus**

1827 publicerade Robert Brown en skildring av små partiklarnas stokastiska rörelse, som man kan betrakta till exempel i en droppe på objektivglaset under ett mikroskop. Sedan dess blev förklaringen till den Brownska rörelsen länge en olöst fråga. Först i början av 1900-talet med Albert Einsteins välkända verk från 1905 och 1908 insågs det att rörelsen har den termiska energin som drivkraft.

*"Termisk agitation av elektronerna i en elektrisk ledning"* kallade John Bertrand Johnson sin upptäckt av termiskt brus i sina mätningar, som han publicerade 1927 [1]. En fysikalisk härledning för fenomenen gavs ett år senare av Harry Nyquist. Einsteins teori om Brownsk rörelse och Nyquists teori om elektriskt brus ledde i början av 1950-talet i statistiska fysiken till fluktuations-dissipations teoremet, som numera är ett generellt verktyg för att beskriva fenomen där dissipation genereras av fluktuationer i ett jämviktstillstånd.

Johnson var född 1887 i Göteborg som ett faderlöst barn, medan Nyquist föddes två år senare i Värmland till föräldrar med sju barn. Båda emigrerade som 17–18 åringar till Amerika, där de senare blev studiekamrater vid Yale universitet. År 1917, efter att ha avlagt sina doktorsavhandlingar i fysik samma år, blev de också arbetskamrater som forskare vid AT&T Bell Telephone Laboratories i New Jersey.

I sitt nya arbete började Johnson studera, med Walter Schottky som förebild, elektriskt brus i ett vakuumrör. Bruset förorsakas av fluktuationer i strömmen av enskilda laddningsbärare som emitteras från katoden. Fenomenen döptes till "shot noise" eller hagelbrus som växer med strömmen. Men när vakuumrörförstärkaren var ansluten till en resistans, då konstaterade Johnson att bruset hade en annan extra komponent som inte var strömberoende. Det visade sig, att medeltalet av kvadraten på spänningsfluktuationerna i detta brus, $\langle V^2 \rangle$, var proportionell till resistansvärdet $R$ och den absoluta temperaturen $T$, men helt oberoende av resistansens material, geometri eller form. Han underströk speciellt, att beroendet av resistansens temperatur betydde att detta brus inte kommer från vakuumrören i förstärkaren, utan att det uppstår i den elektriska resistansen genom statistiska fluktuationer av de elektriska laddningsbärarna. Den termiska rörelsen av ledningselektronerna, även i ett termiskt jämvikts tillstånd, producerar brus. Storleken av bruset var enligt hans mätningar $\sqrt{\langle V^2 \rangle} = 6$ μV inom en bandbredd av $\Delta f$ = 5 kHz över en resistans av $R$ = 500 kΩ vid rumstemperatur.

Johnsons mätningar visade mycket klart vilka ingredienser ingår i termiskt brust och Nyquists härledning gav en tydlig fysikalisk förklaring. Det finns idag ett flertal sätt att bevisa Nyquists uttryck för bruseffekten. Nyquist inledde sitt bevis genom att använda principen om detaljerad balans vid ett termiskt jämviktsförhållande, som Ludwig Boltzmann använde i början av 1870-talet för att beskriva inverkan av partikelkollisionerna i kinetisk gasteori, och utnyttjar Boltzmanns ekvipartitionslag av den termiska energin bland systemets frihetsgrader och därtill Maxwells klassificering av elektromagnetiska egentillstånd, som består av både elektrisk och magnetisk energi. Detta leder till det enkla resultatet för den genomsnittliga kvadratiska brusspänningen $\langle V_T^2 \rangle$,

$$\langle V_T^2 \rangle = 4k_B TR \, \Delta f , \qquad (1)$$

som har visat sig att vara en korrekt beskrivning för den praktiska situationen i vanliga laboratorieförhållanden. Men den innefattar inte existensen av den kvantmekaniska nollpunktsenergin, när $T \to 0$, och inte heller ultraviolett katastrofen, när $T \to \infty$. Den senare hade blivit känt från Rayleigh-Jeans-lagens ofullständiga beskrivning av svartkroppsstrålningen vid höga frekvenser, för vilken Max Planck presenterade den kvantmekaniska lösningen år 1900. Nyquist slutade sitt bevis för (1) genom att hänvisa till Plancks förklaring av ultraviolett katastrofen.

-----------------------------------------------------------

**Skribenterna:**

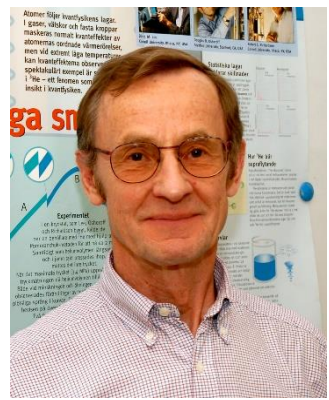

**Matti Krusius** är professor emeritus vid Aalto-universitets institution för teknisk fysik. Hans forskningsområde är lågtemperaturfysik och kryogenik, där hans huvudsakliga intresset är supraledare och speciellt helium superfluider.

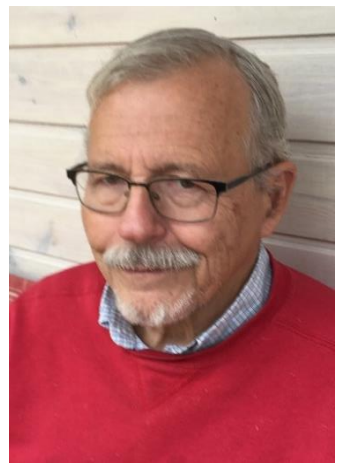

**Gösta Ehnholm** är docent och äldre rådgivare vid Aalto-universitets institution för neurovetenskap och biomedicinsk teknik. Han har forskat i kryogenik och arbetat som industriell utvecklingschef för bl.a. MRI och ultraljudsterapi, senast vid Philips Medical.